\documentclass[conference]{IEEEtran}
\IEEEoverridecommandlockouts
\usepackage[most]{tcolorbox}
\usepackage{amsmath}
\usepackage{amsthm}
\usepackage{url}
\usepackage{tikz}
\usepackage{multirow}

\usepackage{colortbl}

\usepackage{graphicx}
\usepackage{xcolor}
\usepackage{color}

\usepackage{varwidth}
\usepackage{rotating}

\usepackage[none]{hyphenat} 

\usepackage{nameref}
\usepackage{zref-xr,zref-user}
\zxrsetup{toltxlabel=true, tozreflabel=false}
\usepackage{xcite}

\usepackage{tikz-3dplot}

\usepackage{subfigure}

\usetikzlibrary{decorations.pathreplacing,decorations.markings}

\tikzset{arrow data/.style 2 args={%
      decoration={%
         markings,
         mark=at position #1 with \arrow{#2}},
         postaction=decorate}
      }%

\usetikzlibrary{shapes,backgrounds,calc}

\tikzstyle{dummy} = [rectangle, text width=0.1em, draw=white, white,
                      minimum width=0.1em, minimum height=3em, opacity=0.0]

\makeatletter
\newcommand*{\Strut}[1][0.1em]{\vrule\@width\z@\@height#1\@depth\z@\relax}
\makeatother

\begin{document}

\title{Is It Safe Living in the Vicinity of Cellular Towers? \\Analysis of Long-Term Human EMF Exposure \\at Population Scale}
\author{Luca Chiaraviglio,$^{(1,2)}$ Cristian Di Paolo,$^{(1,2)}$ Giuseppe Bianchi,$^{(1,2)}$ Nicola Blefari-Melazzi$^{(1,2)}$\\
(1) Department of Electronic Engineering, University of Rome Tor Vergata, Rome, Italy, \\email \{luca.chiaraviglio,giuseppe.bianchi,blefari\}@uniroma2.it\\
(2) Consorzio Nazionale Interuniversitario per le Telecomunicazioni, Italy
}

\maketitle

\IEEEpeerreviewmaketitle

\begin{abstract}
We focus on the ElectroMagnetic Field (EMF) exposure safety for people living in the vicinity of cellular towers.
To this aim, we analyze a large dataset of long-term EMF measurements collected over almost 20 years in more than 2000 measurement points spread over an Italian region. We evaluate the relationship between EMF exposure and the following factors: (\textit{i}) distance from the closest installation(s), (\textit{ii}) type of EMF sources in the vicinity, (\textit{iii}) Base Station (BS) technology, and (\textit{iv}) EMF regulation updates. Overall, the exposure levels from BSs in the vicinity are below the Italian EMF limits, thus ensuring safety for the population. Moreover, BSs represent the lowest exposure compared to Radio/TV repeaters and other EMF sources. However, the BS EMF exposure in proximity to users exhibits an increasing trend over the last years, which is likely due to the pervasive deployment of multiple technologies and to the EMF regulation updates. As a side consideration, if the EMF levels continue to increase with the current trends, the EMF exposure in proximity to BSs will saturate to the maximum EMF limit by the next 20 years at a distance of 30 meters from the closest BS.
\end{abstract}


\section{Introduction}
\label{sec:intro}
The installation of cellular towers hosting Base Station (BS) functionalities is a fundamental step to provide the variegate services that are required by mobile users.\footnote{In this work, the terms BSs and cellular towers are interchangeable used, since the majority of BSs in Italy are installed either on stand-alone poles or on roof-mounted poles. To the best of our knowledge, the exploitation of BSs not mounted on poles (e.g., micro BSs installed on buildings facades) is very limited in Italy.} Although previous works in the literature  \cite{icnirpnote,roosli2010systematic} demonstrate the lack of proven health effects triggered by living in the vicinity of cellular towers, the debate about public health consequences due to BS ElectroMagnetic Field (EMF) exposure is a controversial aspect among the population. In many countries in the world, the installation of BSs is subject to very stringent constraints, which impose e.g., very low EMF exposure levels from BSs, as well as minimum distances that have to be ensured w.r.t. sensitive places in the vicinity of the installation(s). In Italy, for example, the EMF exposure from cellular towers is subject to a maximum EMF limit set to 6~[V/m] in residential zones \cite{d381-1998}. In addition, a minimum distance of 100 meters between BSs and sensitive places is enforced in many cities (e.g., in Rome \cite{marino}). Although the administrative and legal procedures to authorize the installation of cellular towers are rigorous and clear, a general feeling of fear is shared by the inhabitants living in proximity to BS sites. This fear is exacerbated by many allegations against BS exposure appearing in the social media, which include e.g., the suspect that the installation of BSs is driven by revenue policies and not by public health considerations for the exposed population. 

In this context, a natural question emerges: Is it safe living in the vicinity of cellular towers in terms of health? More concretely, we target the problem of EMF exposure safety by analyzing the long-term EMF exposure levels for the population living in proximity to cellular towers and their positioning w.r.t. the strict EMF limit currently enforced in residential areas of Italy. Although the problem may be clear for the research community actively involved in the topic, in this work we try to shed light on it in a way that is understandable also by researchers working on other topics and more in general by general public.  

Previous works in literature target specific aspects of the problem, which include e.g., the evaluation of the exposure over limited zones of the territory \cite{joseph2012situ} (e.g., single cities), a limited set of targeted mobile technologies \cite{vsuka2019application,koprivica2016statistical} (e.g., only 3G and/or only 4G), and/or measurements performed over a limited amount of time \cite{urbinello2014use,aerts2019situ}. Although we recognize the importance of such previous studies, in this work we go five steps further by: (\textit{i}) considering a very large dataset made available by the Regional Environmental Protection Agency (ARPA) \cite{arpae}, spanning over almost 20 years of measurements that were performed on a vast Italian region, which is covered by multiple operators and by multiple mobile technologies (from 2G up to 4.5G); (\textit{ii}) analyzing how much the measured EMF levels are affected by the distance between the measurement point and the closest BS(s); (\textit{iii}) comparing the BS exposure against other EMF sources (e.g., Radio/TV repeaters) that are installed close to users; (\textit{iv}) evaluating how much the deployment of subsequent BS technologies (e.g., 3.5G, 4G, 4G+) and the EMF regulation updates over the years have impacted the EMF exposure levels in proximity to the BSs; (\textit{v}) investigating the evolution of the EMF exposure for the population living close to BSs during the next 20-30 years, by assuming that the EMF levels will continue to increase with the current trends. 

Our results indicate that the EMF exposure levels in the vicinity of cellular towers is largely below the 6~[V/m] Italian limit, thus providing an adequate safety level for the population. Moreover, we demonstrate that BSs generate a consistently lower amount of exposure compared to other EMF sources (like Radio/TV repeaters). However, the yearly evolution of EMFs reveals that the exposure levels exhibit a slightly increasing trend over the last 4-5 years. This is likely due to the pervasive deployment of new BS technologies (4G/4G+), as well as the modifications that were performed in the Italian EMF regulations about the compliance procedure to verify the adherence to the limits. Eventually, the EMF exposure levels at 30 meters of distance from BSs may saturate to the 6~[V/m] limit by the next 20 years. This condition will increase the fear about BS EMF exposure by the population on one side, and it will severely impact the deployment of new cellular towers by the operator on the other side.

\section{Dataset Description}

Tab.~\ref{tab:area_features} reports the main features of the area taken into account in this study. Specifically, we consider a wide set of measurements\footnote{The raw measurement data are publicly available at \cite{arpae}.} performed over Emilia-Romagna, a vast Italian region, inhabited by more than 4~[million] people. The region is covered by four main mobile operators, which provide 2G/3G/4G/4G+ services (2019 reference year). In this scenario, more than 5000 BSs are actually deployed over the territory. Each BS is a Radio Frequency Source (RFS) for the population living in the vicinity. Clearly, the inhabitants of Emilia Romagna are also exposed to the RFSs generated by legacy Radio/TV repeaters. In addition, other RFSs radiating over the territory include: WiMax equipment, TETRA equipment, train communication equipment, civil/military radars, DVB-H repeaters, S-DAB repeaters, and not-classified RFSs. By analyzing in more detail Tab~\ref{tab:area_features}, we can note that the majority of RFSs is represented by BSs, while the number of Radio/TV RFSs and other RFSs is consistently lower than the one of deployed BSs. This outcome is expected because BSs are pervasively installed over the territory, in order to provide a set of mobile services to users. 

The EMF measurements under consideration are performed inside/outside private buildings, as well as in proximity to sensitive places, such as schools, hospitals, retirement houses, and nursing houses. The locations for performing the measurements have been chosen by ARPA in accordance with the local municipalities, by prioritizing places in proximity to critical installation points (e.g., a location very close to the RFS, or simultaneous presence of multiple RFSs of different types in the neighborhood). In this work, the locations of the measurements are referred as Test Points (TPs). Each measurement is typically carried out by installing at the selected TP an EMF equipment (i.e., a professional wide-band meter), which continuously records the measured EMF with a resolution of 0.01~[V/m]. In addition, the distance between TP and the closest RFS(s), the type of the closest RFS(s) and the duration of the measurements are also recorded. The collected data are then sent at regular intervals (i.e., typically every 24~[h]) to a central processing server managed by ARPA. 

When the measurement is completed, the central processing server computes a set of metrics over the samples recorded during the measurement. The consolidated data include: (\textit{i}) the maximum recorded EMF, (\textit{ii}) the maximum among the 24~[h] average EMF computed for each day of the measurement period, (\textit{iii}) the average EMF computed over the whole measurement period. In this work, we are interested in analyzing the long-term average exposure of EMFs over the population. Therefore, we discard option (\textit{i}), since the RFSs are in general subject to large variations of the radiated EMF over time \cite{vsuka2019application}. As a consequence, the adoption of the maximum EMF over the whole interval would lead to an over-estimation of the actual levels of long-term exposure. In addition, we found that option (\textit{ii}) introduces a slight over-estimation  of the exposure w.r.t. option (\textit{iii}) (in the range 9-15~[\%] on average). This is again due to the fact that the EMF radiated by RFSs may also change across different days. Since our aim is to analyze the exposure levels over long time intervals, we select option (\textit{iii}), i.e., the average EMF computed over the whole measurement period.

\begin{table}[t]
\caption{Main features of the considered area (2019 update)}
\label{tab:area_features}
\centering
\begin{tabular}{|c|c|}
\hline
\textbf{Feature} & \textbf{Value} \\
\hline
Region name & Emilia-Romagna \\
Area of interest & 22453 km$^2$\\
Number of inhabitants & $4.59\times10^6$\\
Mobile technologies & 2G/3G/4G/4G+\\
Mobile Operators & 4 (TIM, Vodafone, Wind Tre, Iliad) \\
Number of BS RFSs & 5301 \\
Number of Radio/TV RFSs &  1264\\
Number of Other RFSs & 576 \\
\hline
\end{tabular}
\vspace{-4mm}
\end{table}

We now provide more details about the metrics stored in the whole dataset. Let us denote with $\mathcal{N}$ the set of measurements under consideration. The set of possible RFS types is denoted with $\mathcal{L}$. The RFS types in $\mathcal{L}$ are $\{\text{BS},\text{Radio},\text{TV},\text{Other}\}$. The set of years is denoted with $\mathcal{Y}$. Each measurement $n \in \mathcal{N}$ is then characterized by:
\begin{itemize}
\item measurement duration $t_n$ (in days);
\item average EMF $f_n$ (in Volt per meter) computed over $t_n$;
\item distance $d_n$ (in meters) between the TP and the closest RFS(s);
\item ending year $y_n \in \mathcal{Y}$ of the measurement;
\item set of binary parameters $s_{(n,l)}$, $\forall l \in \mathcal{L}$. Each parameter $s_{(n,l)}$ takes value 1 if there is at least one RFS of type $l$ is in the vicinity of the TP where measurement $n$ has been performed,\footnote{The vicinity between one RFS and one TP is defined if the RFS is in Line-of-Sight conditions w.r.t. the building hosting the TP. For cellular towers, such distance is typically lower than 700~[m] in the considered dataset.} 0 otherwise.
\end{itemize}
Focusing on the $f_n$ values, when the average EMF over $t_n$ is lower than $0.5$~[V/m], the string ``$<0.5$'' is written in record $f_n$. Otherwise, if the average EMF is higher than 0.5~[V/m], $f_n$ reports the value of the average EMF level. The reason for not recording the exact value when the EMF level is lower than $0.5$~[V/m] is double. On one side, in fact, such EMF levels are so low that the presence of interfering sources (like mobile terminals) in the TP proximity may lead to large errors in the assessment of the EMF from the BSs. On the other side, the equipment used to measure the EMF levels may also require a minimum EMF level to provide a reliable result. 

\begin{table}[t]
\caption{Measurements Description}
\label{tab:dataset_features}
\centering
\begin{tabular}{|c|c|c|}
\hline
\textbf{Measurement Feature} & \textbf{Notation} & \textbf{Value(s)} \\
\hline
Number of measured RFSs & $|\mathcal{N}|$ &2699 \\
Number of measured RFSs (filtered) & $|\mathcal{N}^{\text{F}}|$ & 2410  \\
Number of B measurements & $|\mathcal{N}^{\text{B}}|$ &1990\\
Number of (R,T) measurements & $|\mathcal{N}^{\text{R,T}}|$ & 188 \\
Number of BR/T measurements & $|\mathcal{N}^{\text{BR/T}}|$  & 66 \\
Number of BRT measurements & $|\mathcal{N}^{\text{BRT}}|$ & 148 \\
Number of O measurements & $|\mathcal{N}^{\text{O}}|$ & 18 \\
Minimum measurement duration & $\min_n (t_n)$ & 1 day\\ 
Covered years & $\mathcal{Y}$ & 2003-2019 \\
\hline
\end{tabular}
\vspace{-4mm}
\end{table}

Tab.~\ref{tab:dataset_features} provides more details about the values stored in the dataset. The total number of measurements $|\mathcal{N}|$ is roughly equal to 2700. Given $\mathcal{N}$, we discard the records reporting missing information. Specifically, a measurement $n \in \mathcal{N}$ is discarded if at least one of the following conditions is met: (\textit{i}) missing duration $t_n$, (\textit{ii}) missing EMF level $f_n$, (\textit{iii}) missing RFS-TP distance $d_n$, (\textit{iv}) missing RFS, i.e., $\sum_{l \in \mathcal{L}} s_{(n,l)}==0$. The resulting filtered set is denoted as $\mathcal{N}^F$. Interestingly, as reported in Tab.~\ref{tab:dataset_features}, the cardinality of $\mathcal{N}^F$ is still very large (i.e., more than 2400 measurements). We then extract from $\mathcal{N}^F$ the following categories:
\begin{itemize}
\item \textbf{Only BSs (B)}, i.e., measurements whose closest installation(s) are solely BSs; this subset is formally denoted as: $n \in \mathcal{N}^{\text{B}} \subset \mathcal{N}^{\text{F}}: (s_{(n,l_1)}=1, s_{(n,l_2)}=0), l_1 = \{\text{BS}\}, l_2 =\{\text{Radio},\text{TV},\text{Other}\}$  
\item \textbf{Radio, TV (R,T)}, i.e., measurements from either Radio, TV, or both Radio and TV RFSs (but not BSs and/or other RFSs); this subset is formally expressed as: $n \in \mathcal{N}^{\text{R,T}} \subset \mathcal{N}^{\text{F}}: (s_{(n,l_1)}=1, s_{(n,l_2)}=1, s_{(n,l_3)}=0), [(l_1 = \{\text{Radio}\} \wedge l_2 = \{\text{TV}\}) \vee (l_1,l_2 = \{\text{Radio}\}) \vee (l_1,l_2 = \{\text{TV}\})] \wedge l_3 =\{\text{BS},\text{Other}\}$;
\item \textbf{BSs + Radio/TV (BR/T)}, i.e., measurements from BSs and Radio (or TV) RFSs (but not other RFSs); we denote this subset as: $n \in \mathcal{N}^{\text{BR/T}} \subset \mathcal{N}^{\text{F}}: (s_{(n,l_1)}=1, s_{(n,l_2)}=1, s_{(n,l_3)}=1, s_{(n,l_4)}=0), [l_1 = \{\text{BS}\} \wedge (l_2 = \{\text{Radio}\} \vee l_3 = \{\text{TV}\}) \wedge l_4=\{\text{Other}\}]$;
\item \textbf{BSs + Radio + TV (BRT)}, i.e., measurements from BSs, Radio and TV RFSs (but not other RFSs); this subset is formally denoted as: $n \in \mathcal{N}^{\text{BRT}} \subset \mathcal{N}^{\text{F}}: (s_{(n,l_1)}=1, s_{(n,l_2)}=1, s_{(n,l_3)}=1, s_{(n,l_4)}=0), (l_1 = \{\text{BS}\} \wedge l_2 = \{\text{Radio}\} \wedge l_3 = \{\text{TV}\} \wedge l_4 \{\text{Other})$;
\item \textbf{Other (O)}, i.e., measurements where at least one of closest installations is generated by other RFSs; this subset is formally expressed as: $n \in \mathcal{N}^{\text{O}} \subset \mathcal{N}^{\text{F}}: s_{(n,l)}=1, l = \{\text{Other}\}$. 
\end{itemize}
The cardinality of each subset $\mathcal{N}^{\text{B}}$, $\mathcal{N}^{\text{R,T}}$, $\mathcal{N}^{\text{BR/T}}$, $\mathcal{N}^{\text{BRT}}$ and $\mathcal{N}^{\text{O}}$ is reported in Tab.~\ref{tab:dataset_features}. Interestingly, we can note that the largest set is $\mathcal{N}^{\text{B}}$ (as expected), while all the other subsets exhibit lower cardinalities. This result is somehow expected, due to the following main reasons: (\textit{i}) BSs are pervasively deployed over the territory (as reported in Tab.~\ref{tab:area_features}), (\textit{ii}) Radio/TV repeaters tend to be more sparsely installed over the territory w.r.t. BSs, (\textit{iii}) Radio/TV RFSs are in general installed in less densely populated zones (e.g., on top of the hills), (\textit{iv}) BSs are the major source of fear for the population, resulting in a selection of the TPs polarized towards this category by the local municipalities.

\begin{figure}[t]
\centering
\includegraphics[width=7cm]{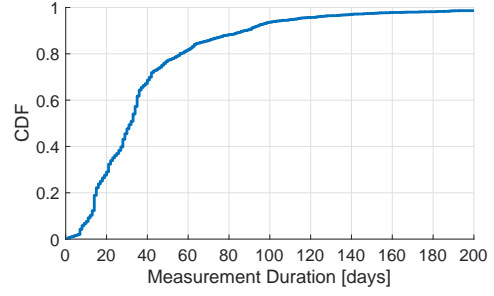}
\label{fig:cdf_duration}
\caption{CDF of the duration of the measurements ($\mathcal{N}^{\text{F}}$ filtered set).}
\vspace{-4mm}
\end{figure}

\begin{figure}[t]
\centering
\includegraphics[width=7cm]{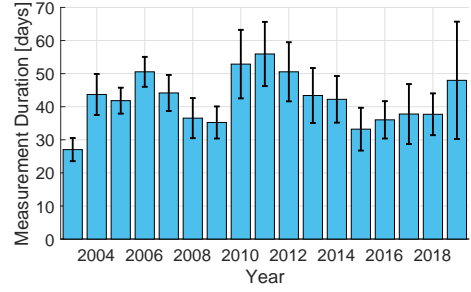}
\label{fig:cdf_duration_time}
\caption{Average measurement duration vs. year ($\mathcal{N}^{\text{F}}$ filtered set).}
\vspace{-4mm}
\end{figure}

Eventually, the last two rows of Tab.~\ref{tab:dataset_features} report the minimum duration for each measurement (equal to one day), and the covered years in the dataset, i.e., equal to 17 years including starting and ending year. A natural question is then: How does the duration vary across the measurements and across the years? To this aim, Fig.~\ref{fig:cdf_duration} reports the Cumulative Distribution Function (CDF) of $t_n$ in $\mathcal{N}^{\text{F}}$. Interestingly, while the minimum measurement duration is equal to one day, the majority of the measurements have been performed over a long time-scale, i.e., $t_n$ is  typically in the order of dozens of days. This is beneficial for a correct evaluation of the long-term exposure by the population. In addition, Fig.~\ref{fig:cdf_duration_time} reports the evolution of $t_n$ over the years. Bars report the average values of $t_n$, while error bars report the confidence intervals (computed with a 95\% of confidence level). While variations of $t_n$ are experienced across the years (see e.g., 2010 w.r.t. 2009), the average duration is always higher than 20 days. Therefore, we can claim that the considered dataset allows a fair evaluation of the long-term exposure across the years.

Focusing on the EMF levels, we proceed as follows. For the measurements whose $f_n$ field is labelled with ``$<0.5$'', we either impose $f_n=0.1$~[V/m] or $f_n=0.5$~[V/m]. We denote the two alternatives as $F_{\text{MIN}}$=0.1~[V/m] and $F_{\text{MIN}}$=0.5~[V/m], respectively. The setting $F_{\text{MIN}}$=0.1~[V/m] assumes that all the measurements labelled with ``$<0.5$'' are all very low, i.e., close to a negligible EMF. On the other hand, the setting $F_{\text{MIN}}$=0.5~[V/m] introduces a very conservative assumption in terms of EMF, i.e., all the measurements labelled with ``$<0.5$''  are close to 0.5~[V/m]. In the rest of the work, we will provide the results by selectively activating the two options, i.e., either $F_{\text{MIN}}$=0.1~[V/m] or $F_{\text{MIN}}$=0.5~[V/m].


\begin{figure}[t]
\centering
\subfigure[$F_{\text{MIN}}$=0.1~V/m]
{
	\includegraphics[width=7cm]{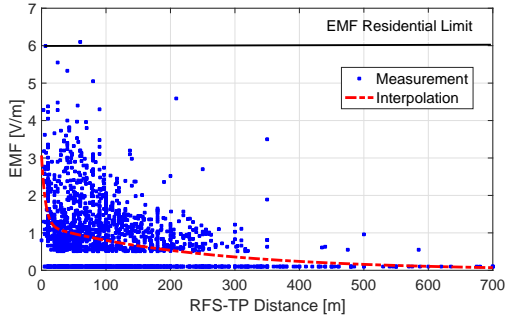}
	\label{fig:c_vs_distance_0_1}
}
\subfigure[$F_{\text{MIN}}$=0.5~V/m]
{
	\includegraphics[width=7cm]{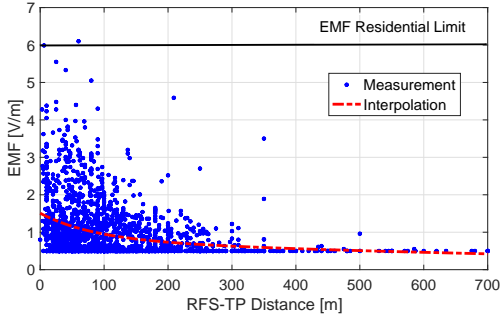}
	\label{fig:c_vs_distance_0_5}
}
\caption{EMF vs. RFS-TP distance (B category).}
\label{fig:c_vs_distance}
\vspace{-4mm}
\end{figure}

\section{Long-Term Analysis of Human Exposure}

We divide our analysis in the following branches: (\textit{i}) investigation of distance on the total exposure, (\textit{ii}) impact of the RFS type, (\textit{iii}) investigation of the impact of BS deployment and EMF regulations on the exposure and (\textit{iv}) evolution of EMF exposure levels.

\subsection{Impact of the Distance from Cellular Towers}

We initially consider the Only BSs (B) category. Fig.~\ref{fig:c_vs_distance_0_1}-\ref{fig:c_vs_distance_0_5} report the EMF vs. the distance for  $F_{\text{MIN}}$=0.1~[V/m] and $F_{\text{MIN}}$=0.5~[V/m], respectively.  Each point is a measurement value, while the dashed lines mark the best interpolation, obtained with a double exponential function,\footnote{Different interpolations (linear, quadratic, single exponential, double exponential) have been tested. The best one is chosen in accordance to the lowest Root Mean Square Error (RMSE).} formally expressed as follows:
\begin{equation}
\label{eq:double:exp}
F_d = \alpha \cdot \text{exp}({\beta \cdot d}) + \gamma \cdot \text{exp}({\delta \cdot d})
\end{equation}
where $d$ is the distance, while $\alpha$, $\beta$, $\gamma$, and $\delta$ are input parameters (shown in Tab.~\ref{tab:estimated_parameters}). Finally, the 6~[V/m] limit currently enforced in Italy in  residential areas is marked with a horizontal line on the top of the figures.

Several considerations hold for Fig.\ref{fig:c_vs_distance_0_1}-\ref{fig:c_vs_distance_0_5}. First, the measured EMF $f_n$ is typically lower than 6~[V/m], in accordance to the maximum EMF limit enforced by law.\footnote{For the measurements above the EMF limit, the operators are obliged to apply a scaling factor to the radiated power of the BS(s), in order to meet the limit.} Second, $f_n$ is rapidly decreasing with the distance (as expected). Third, a huge variability in the measurements is observed for TPs in proximity to the BSs (left part of the figure). This result is meaningful, since the sight conditions w.r.t the closest BS(s) (e.g., Line-Of-Sight or Non-Line-of-Sight) strongly affect the measured EMF levels. Fourth, the interpolated lines reveal that, on average, the EMF levels are already lower than 1~[V/m] when the RFS-TP distance is larger than 100~[m]. Fifth, with the $F_{\text{MIN}}$=0.1~[V/m] assumption, the EMF levels are close to 0~[V/m] for users living at a RFS-TP distance of more than 500~[m]. Clearly, when $F_{\text{MIN}}$=0.5~[V/m], the minimum EMF level is equal to 0.5~[V/m] even for large distances from the RFS. In any case, however, this value is clearly lower than the residential limit of 6~[V/m], thus confirming the safety level for the population.

\begin{table}[t]
\centering
\caption{Estimated parameters for the EMF vs. distance}
\label{tab:estimated_parameters}
\begin{tabular}{|c|c|c|c|c|}
\hline
$F_{\text{MIN}}$ & $\alpha$  & $\beta$ & $\gamma$ & $\delta$ \\
\hline
0.1~[V/m]  & 1.869 & -0.1667 & 1.2 & -0.004044 \\
\hline
0.5~[V/m] & 0.7113 & -0.01117 & 0.7973 &  -0.0009069 \\
\hline
\end{tabular}
\vspace{-3mm}
\end{table}

\subsection{Impact of Nearby Radio-Frequency Sources}

\begin{figure}[t]
\centering
\subfigure[Average EMF ($F_{\text{MIN}}$=0.5~V/m case)]
{
    \includegraphics[width=6.8cm]{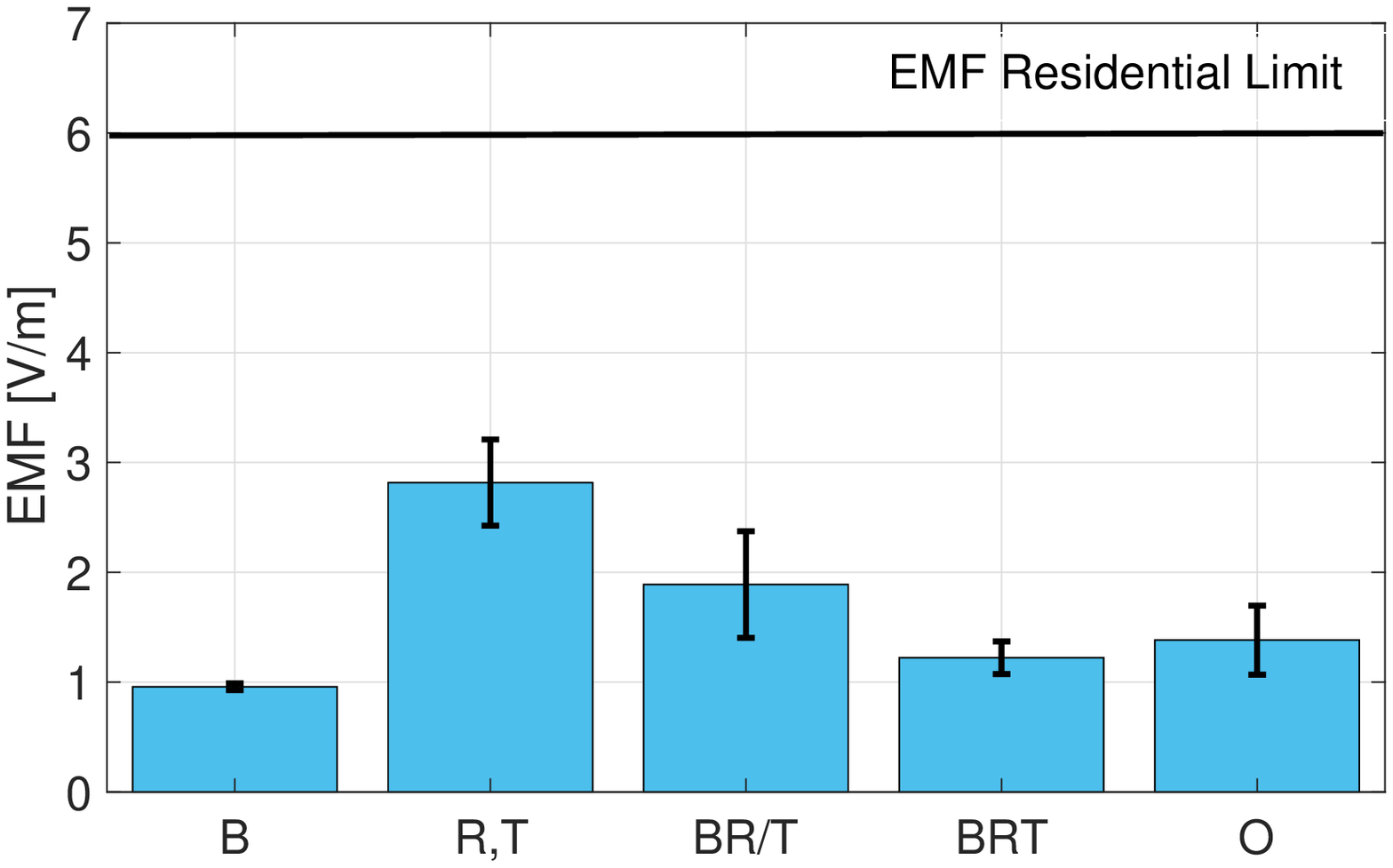}
    \label{fig:f_cat_0_5}
}

\subfigure[CDFs of the EMF ($F_{\text{MIN}}$=0.5~V/m case)]
{
	\includegraphics[width=7cm]{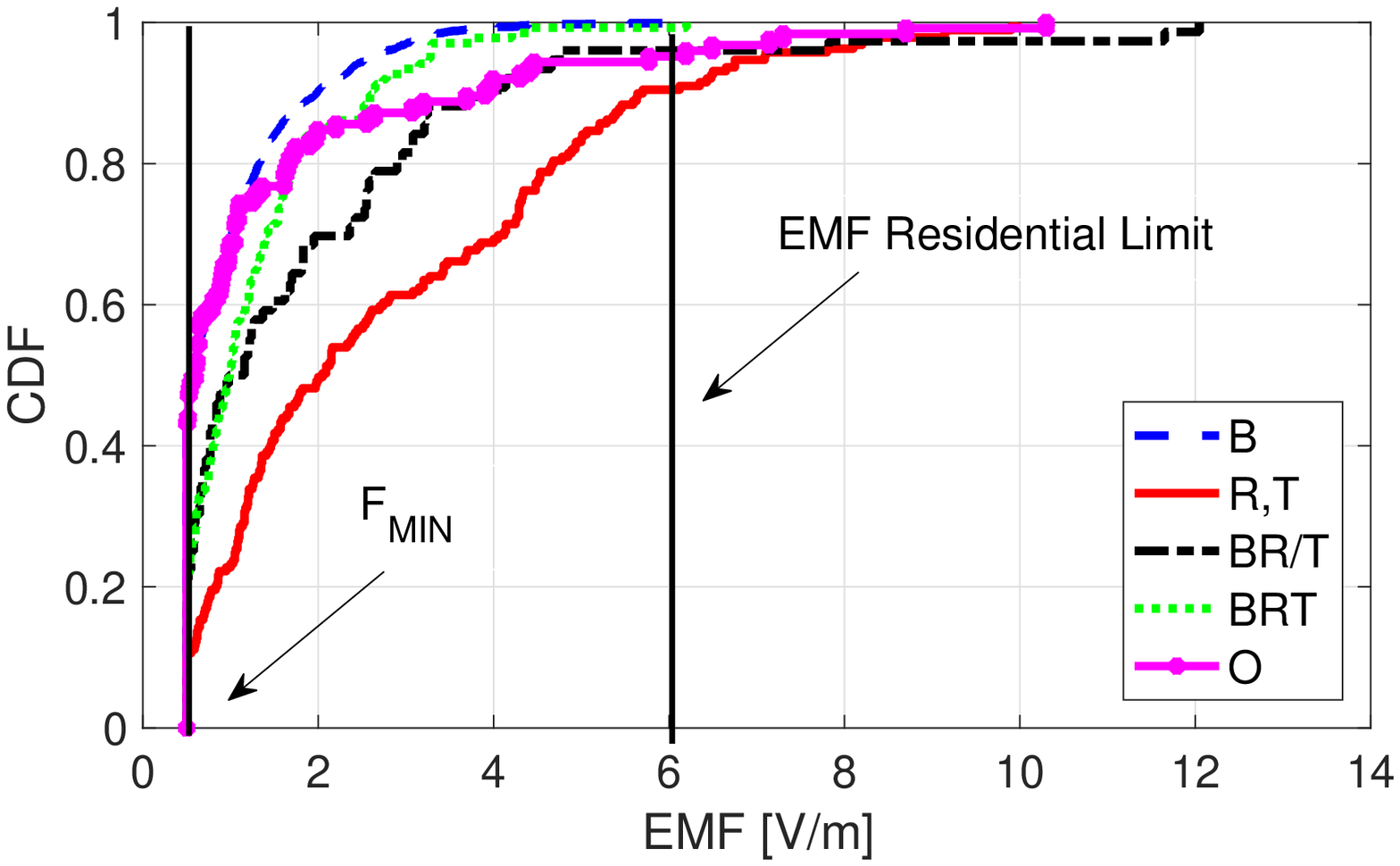}
	\label{fig:cdf_cat_0_5}
}

\subfigure[CDFs of the RFS-TP Distance]
{
	\includegraphics[width=6.9cm]{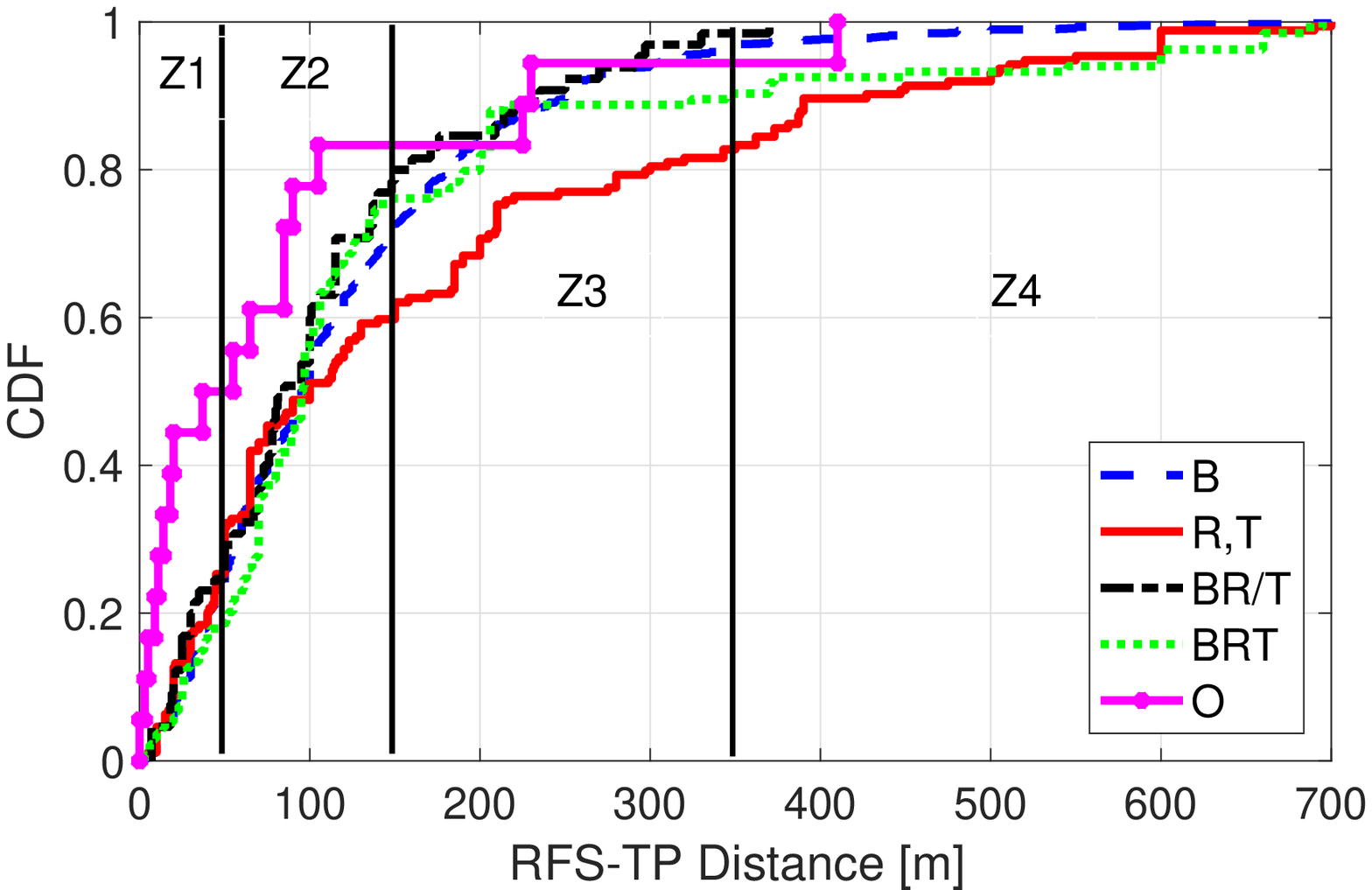}
	\label{fig:cdf_distance_cat:0_5}
}
\caption{Comparison across the different categories in terms of: (\textit{i}) average EMF, (\textit{ii}) CDF of the EMF, (\textit{iii}) CDF of the RFS-TP distance}
\label{fig:cat_comp}
\vspace{-3mm}
\end{figure}

We then compare the EMF radiated from the B category w.r.t the (R,T), BR/T, BRT and O ones. We consider in this step the most conservative assumption in terms of minimum EMF, i.e., $F_{\text{MIN}}$=0.5~[V/m].\footnote{The analysis with $F_{\text{MIN}}$=0.1~[V/m] is omitted due to the lack of space. In any case, results are similar to the $F_{\text{MIN}}$=0.5~[V/m] case described in this work.} Fig.~\ref{fig:f_cat_0_5} reports the average EMF levels and the confidence intervals (with 95\% of confidence level) for each category. Interestingly, when the RFSs are solely BSs (first bar in the figure from the left), the average EMF level is consistently lower than the one from BR/T, BRT and O  categories. On the other hand, Radio and TV RFSs represent a major source of EMF radiation for the population living in the vicinity, with an average EMF close to 3~[V/m] (second bar). 

In the following, we move our attention from the EMF average values to the whole range of measurements. To this aim, we evaluate the CDF of the EMF for the different categories. Fig.~\ref{fig:cdf_cat_0_5} shows the obtained results. The measured EMF ranges between 0.5~[V/m] (which corresponds to the $F_{\text{MIN}}$ threshold)  and around 12~[V/m] (which is recorded for the BR/T category). Interestingly, we can note that the CDF of B category clearly lies on the left of the figure compared to the CDFs of (R,T), BR/T, BRT and O categories. Therefore, when the closest installations are solely BSs, the EMF levels are the lowest ones compared to all the other categories.

However, a key question arises: Does the RFS-TP distance have an impact when comparing the different categories? To answer this question, Fig.~\ref{fig:cdf_distance_cat:0_5} reports the CDFs of the RFS-TP distance. Since the CDF of B is clearly on the left compared to the (R,T) one, the TPs of the former are located at a shorter distance from the RFSs compared to the latter. Despite this fact, however, the EMF from BS RFSs is clearly lower compared to Radio/TV one (i.e., see the corresponding CDFs in Fig.~\ref{fig:cdf_cat_0_5}). On the other hand, the CDFs of B, BR/T and BRT all lie in the same region, i.e., the EMFs are measured under similar RFS-TP conditions. Eventually, the CDF of the O category is on the left w.r.t. the B one, i.e., the measurement distance of the former is shorter than the latter. However, we also remind that the number of measurements for the O category is less than 20, while the number of measurement for the B, (R,F), BR/T, BRT categories is equal to 1990, 188, 66 and 148, respectively. Therefore, the number of O samples may be too small to generalize the findings about the measured exposure levels for this category. 

During the last part of this phase, we have considered four ranges of RFS-TP distances, reported in Fig.~\ref{fig:cdf_distance_cat:0_5}, namely:  Z1) $d_n < 50$~[m], Z2) $ 50 \leq d_n < 150 \text{[m] }$, Z3) $ 150 \leq d_n < 350 \text{[m]}$, Z4) $d_n \geq 350$~[m]. We have then computed the average EMF in each range, observing that the B category achieves the lowest average EMF compared to (R,T), BR/T, and BRT.\footnote{We omit the figures showing the average EMF in each range due to the lack of space.}


\subsection{Impact of BS Technologies and EMF Regulations}

\begin{figure}[t]
\centering
\subfigure[EMF vs. year ($F_{\text{MIN}}$=0.1~V/m case)]
{
    \includegraphics[width=7cm]{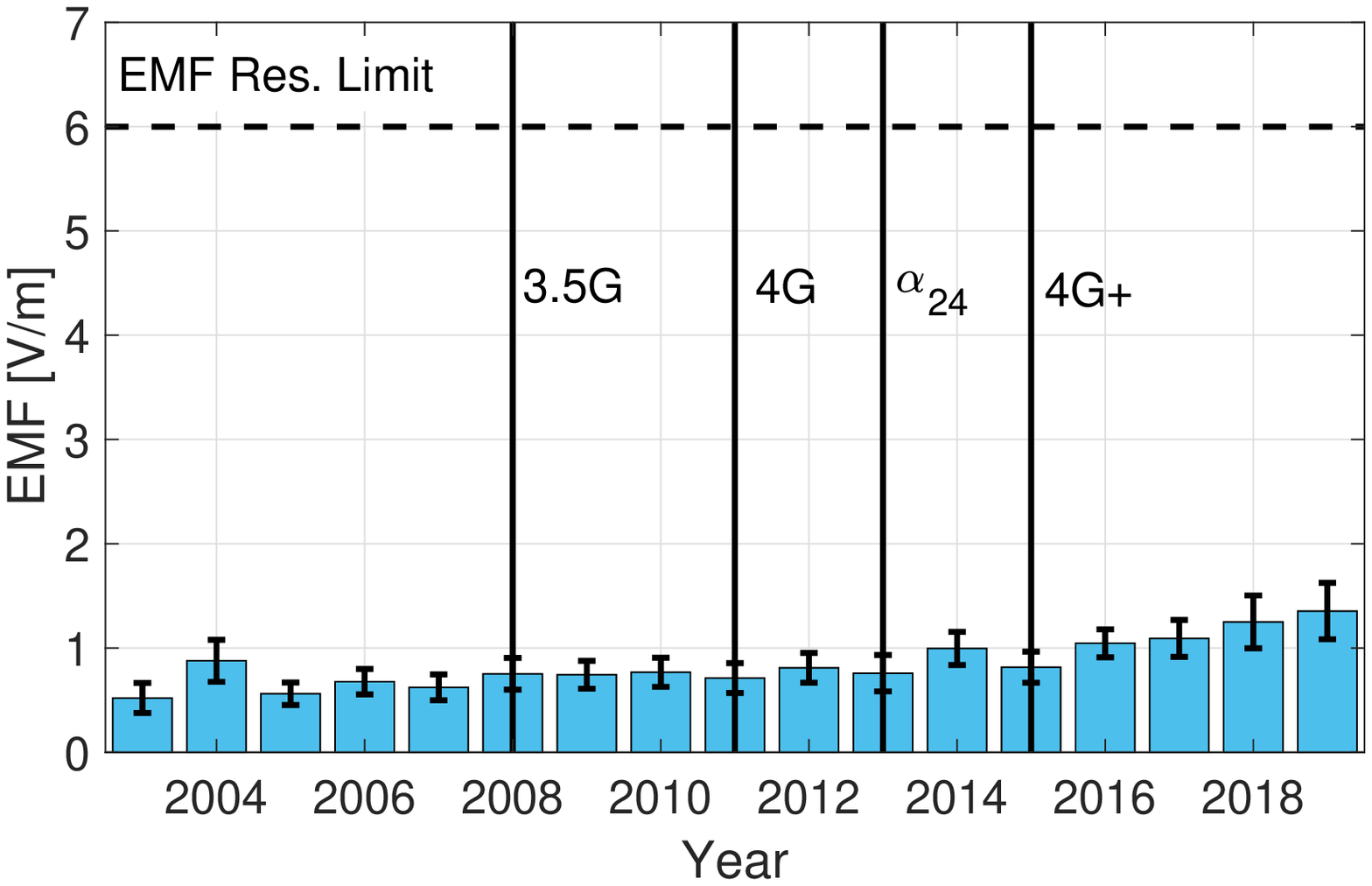}
    \label{fig:f_years_0_1}
}
\subfigure[EMF vs. year ($F_{\text{MIN}}$=0.5~V/m case)]
{
	\includegraphics[width=7cm]{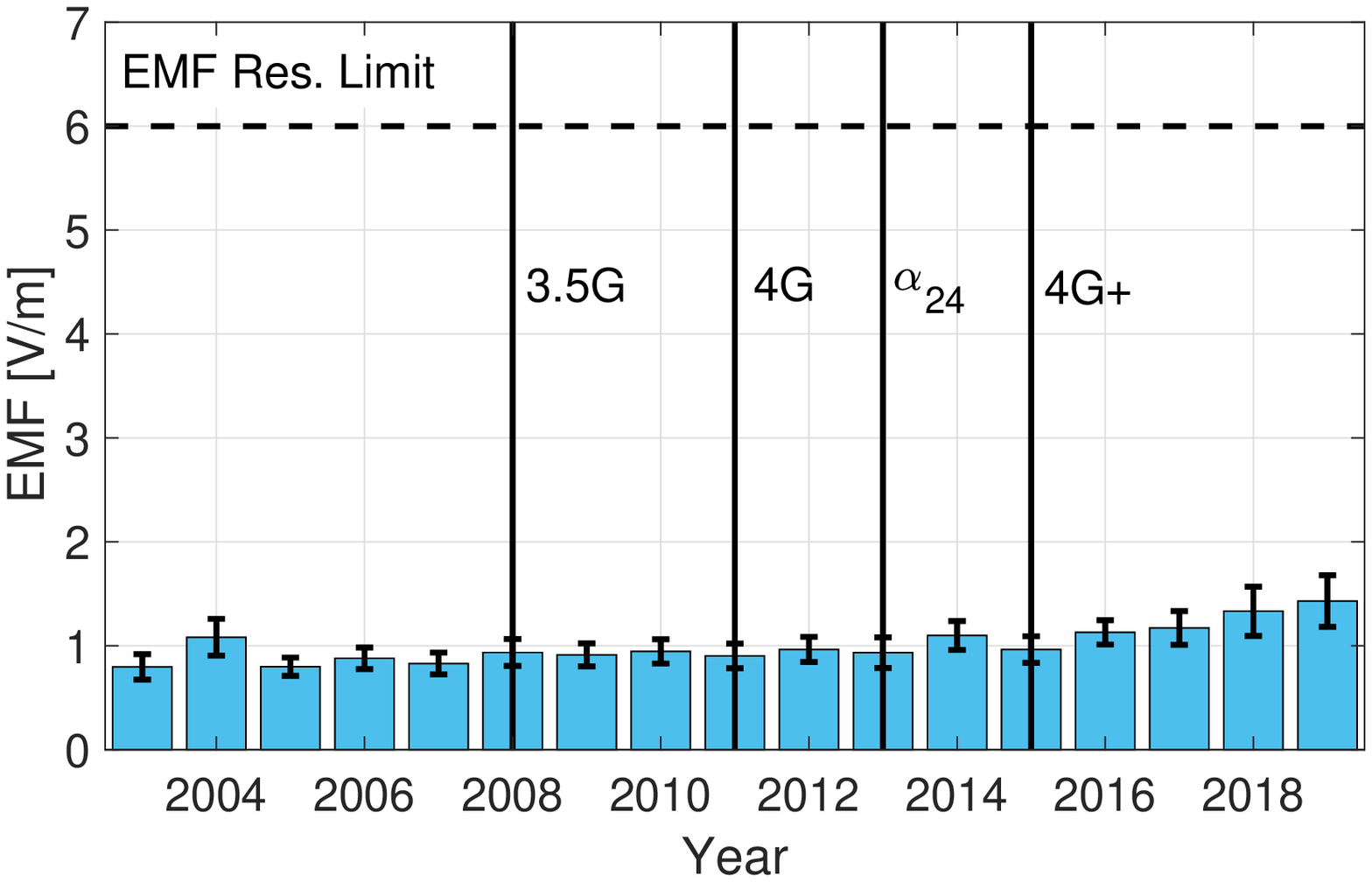}
	\label{fig:f_years_0_5}
}

\subfigure[RFS-TP distance vs. year]
{
	\includegraphics[width=4.1cm]{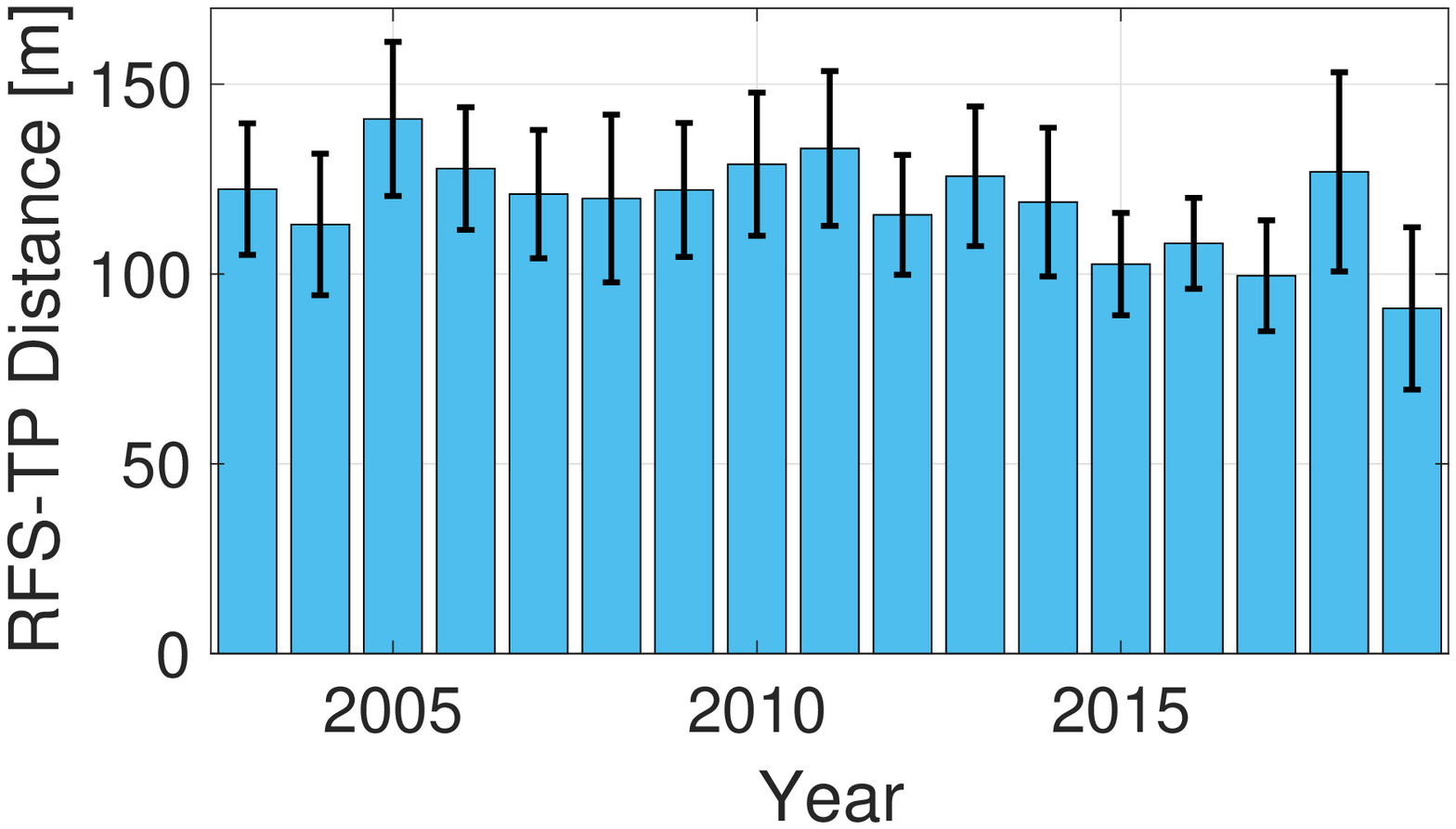}
	\label{fig:distance_years}
}
\subfigure[Number of measurements vs. year]
{
	\includegraphics[width=4cm]{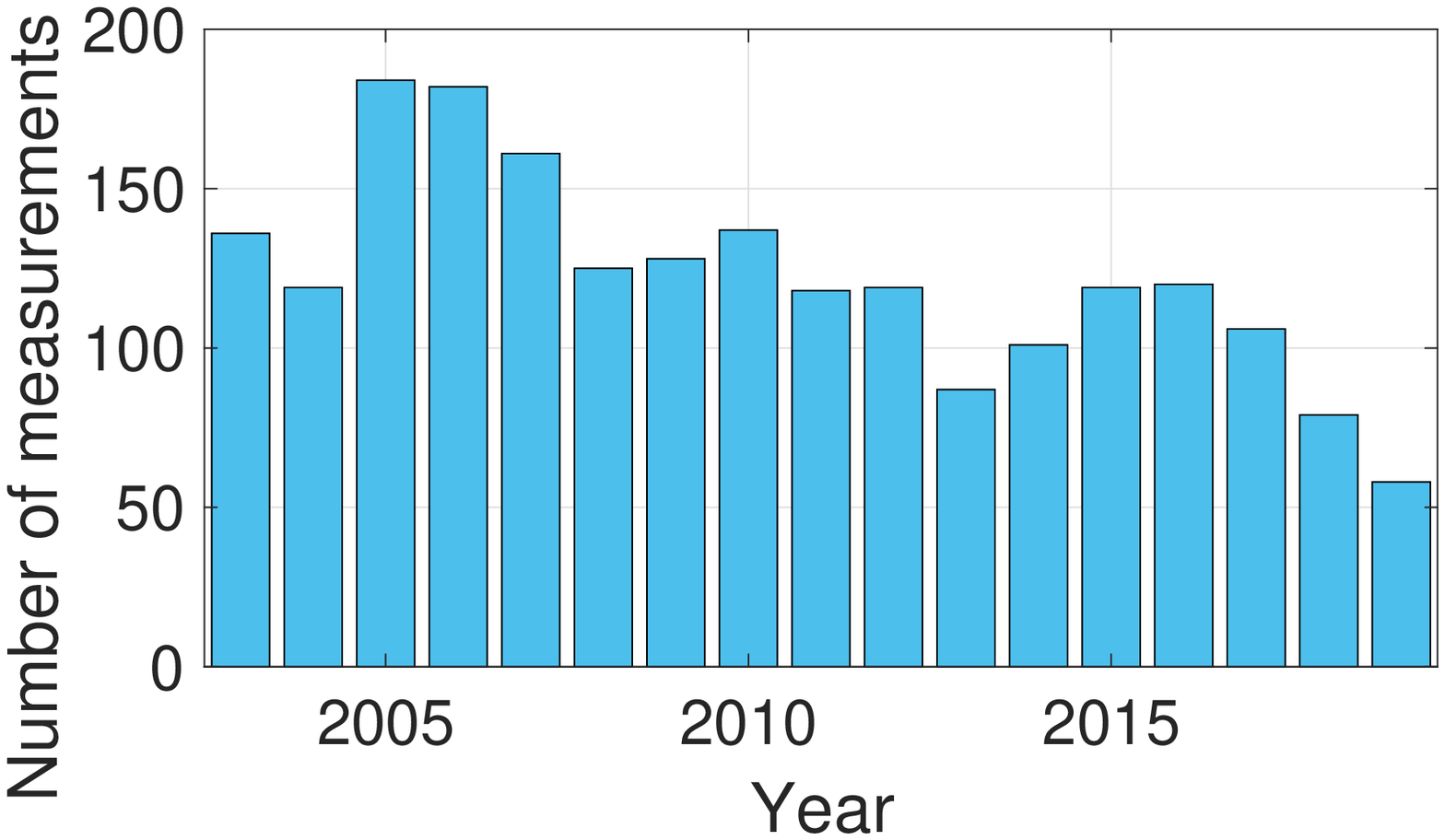}
	\label{fig:n_measurements_years}
}
\caption{Temporal variation of EMF, RFS-TP distance and number of measurements (B category).}
\label{fig:year_var}
\vspace{-3mm}
\end{figure}

We then analyze the temporal evolution of the measurement metrics for the B category, in order to assess the impact of the BS technology deployment and the EMF regulation updates which were performed across the years. 
We initially focus on the evolution of EMFs vs. year for $F_\text{MIN}=0.1$~[V/m] and $F_\text{MIN}=0.5$~[V/m], shown in Fig.~\ref{fig:f_years_0_1}-\ref{fig:f_years_0_5}. The figures report also with a dashed horizontal line the residential EMF limit and with continuous vertical lines the adoption year for the 3.5G/4G/4G+ technologies. In addition, the vertical line $\alpha_{24}$ marks the regulation change \cite{regchange} that introduced less conservative assumptions than the ones previously adopted during the authorization phase when installing new BSs. Specifically, the regulator introduced the parameter $\alpha_{24} \in (0,1]$ to take into account the variation of the BS radiated power, in accordance to the managed traffic and/or number of served users over 24 hours (i.e., higher during the day and lower during the night). The maximum radiated power is then scaled by $\alpha_{24}$ to retrieve the 24h average radiated power. This metric is then used to compute the predicted BS EMF level, which is finally compared to the 6~[V/m] EMF limit. Clearly, when $\alpha_{24} \ll 1$, the 6~[V/m] EMF limit can be more easily satisfied compared to the maximum power case, i.e., $\alpha_{24}=1$. As a consequence, when multiple RFSs sources already contribute to the total EMF level, the introduction of the $\alpha_{24}$ parameter may facilitate the installation of new BSs in residential areas. However, we stress the fact that the $\alpha_{24}$ parameter introduces a realistic assumption (i.e., the BS power variation over time) when computing the radiated EMFs.

\begin{figure}[t]
\centering
\subfigure[$F_{\text{MIN}}$=0.1~V/m]
{
    \includegraphics[width=6.2cm]{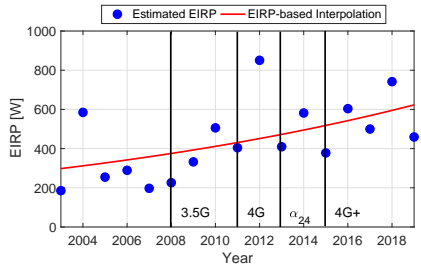}
    \label{fig:eirp_var_0_1}
}
\subfigure[$F_{\text{MIN}}$=0.5~V/m]
{
	\includegraphics[width=6.2cm]{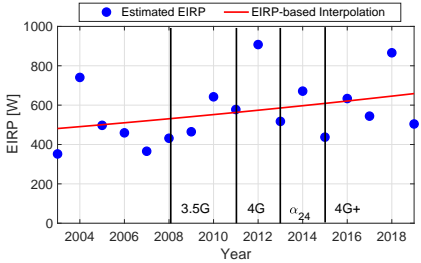}
	\label{fig:eirp_var_0_5}
}
\caption{Estimated EIRP and interpolated EIRP vs. year (B category).}
\label{fig:eirp_var}
\vspace{-3mm}
\end{figure}


%

By observing more carefully Fig.~\ref{fig:f_years_0_1}-\ref{fig:f_years_0_5}, we can note that the EMF levels are clearly lower than the maximum limit. However, the average EMF levels exhibit an increasing trend over the years, either by assuming $F_\text{MIN}=0.1$~[V/m] or $F_\text{MIN}=0.5$~[V/m]. The EMF increase is more evident during the last years (i.e., since 2014). We argue that this increase is likely due to the deployment of multiple BS functionalities (e.g., 3.5G/4G/4G+), together with the application of the $\alpha_{24}$ regulation. To complement these findings, Fig.~\ref{fig:distance_years}-\ref{fig:n_measurements_years} report the RFS-TP distance and the number of measurements vs. the years, respectively. Interestingly, we can note that the RFS-TP distance is almost decreasing over the 2011-2019 years (except from 2018). This finding is in accordance with the pervasive deployment of 4G networks, which have been installed since 2011. On the other hand, the change in the RFS-TP distance may be also explained by a selection of the TPs more polarized towards the ones in close proximity to the RFSs. In any case, the number of measurements (shown in Fig.~\ref{fig:n_measurements_years}) is always pretty large, i.e., always higher than 50 per year.

In order to provide more insights about the observed EMF increase, we compute the Equivalent Isotropic Radiated Power (EIRP) based on the model of \cite{itutk91}, by adopting the following conservative assumptions: (\textit{i}) the RFS is a point source, (\textit{ii}) the closest installation is the only source of EMF exposure, (\textit{iii}) the quadratic decay exponent for the distance is adopted. The motivations of adopting the EIRP are double. On one side, we compute a metric that integrates both EMF and distance. On the other hand, we are able in this way to highlight the evolution of the radiated power over time. Given each measurement $n \in \mathcal{N}^{\text{B}}$, the RFS EIRP is formally expressed as: 
\begin{equation}
\label{eq:eirp_definition}
e_n=\frac{4 \pi \cdot d_n^{(2)} \cdot f_n^{(2)}}{G \cdot Z}
\end{equation}
where $d_n$ is the RFS-TP distance from the closest installation(s), $f_n$ is the measured EMF level, $G=0.96$ is the estimated antenna gain (set in accordance to \cite{itutk91}), and $Z=377$~[$\Omega$]. In the following step, we compute the average EIRP for each year, shown in Fig.~\ref{fig:eirp_var_0_1}-\ref{fig:eirp_var_0_5} for each $F_{\text{MIN}}$ option. The two subfigures also report the interpolated EIRP, which is expressed by the following single exponential function:
\begin{equation}
\label{eq:single:exp}
E_y = \alpha \cdot \text{exp}({\beta \cdot y}) 
\end{equation}
 where $y$ is the year, while $\alpha$ and $\beta$ values are reported in Tab.~\ref{tab:estimated_parameters_EIRP}. By analyzing in more detail Fig.~\ref{fig:eirp_var_0_1}-\ref{fig:eirp_var_0_5}, we can note that the estimated EIRP exhibits an increasing trend over the years, although large oscillations also emerge. In any case, however, the estimated EIRP values confirm that the EMF exposure is increasing over the years. Moreover, this increase is not necessarily due to a decrease of the RFS-TP distance, but rather to the deployment of newer technologies, as well as the introduction of the $\alpha_{24}$ parameter, as we argued when commenting Fig.~\ref{fig:f_years_0_1}-\ref{fig:f_years_0_5}. Eventually, the increasing trend is more evident with $F_{\text{MIN}}=0.1$~[V/m] compared to $F_{\text{MIN}}=0.5$~[V/m].

\begin{table}[t]
\centering
\caption{Estimated parameters for the EIRP vs. year}
\label{tab:estimated_parameters_EIRP}
\begin{tabular}{|c|c|c|}
\hline
$F_{\text{MIN}}$ & $\alpha$  & $\beta$ \\
\hline
0.1~[V/m]  & $2.39 \cdot 10^{-38}$ & 0.0461\\
\hline
0.5~[V/m] &  $4.1381 \cdot 10^{-15}$ & 0.0196 \\
\hline
\end{tabular}
\vspace{-1mm}
\end{table}







\subsection{Evolution of EMF Exposure Levels}
\begin{figure}[t]
\centering
\subfigure[$F_{\text{MIN}}$=0.1~V/m]
{
    \includegraphics[width=6.5cm]{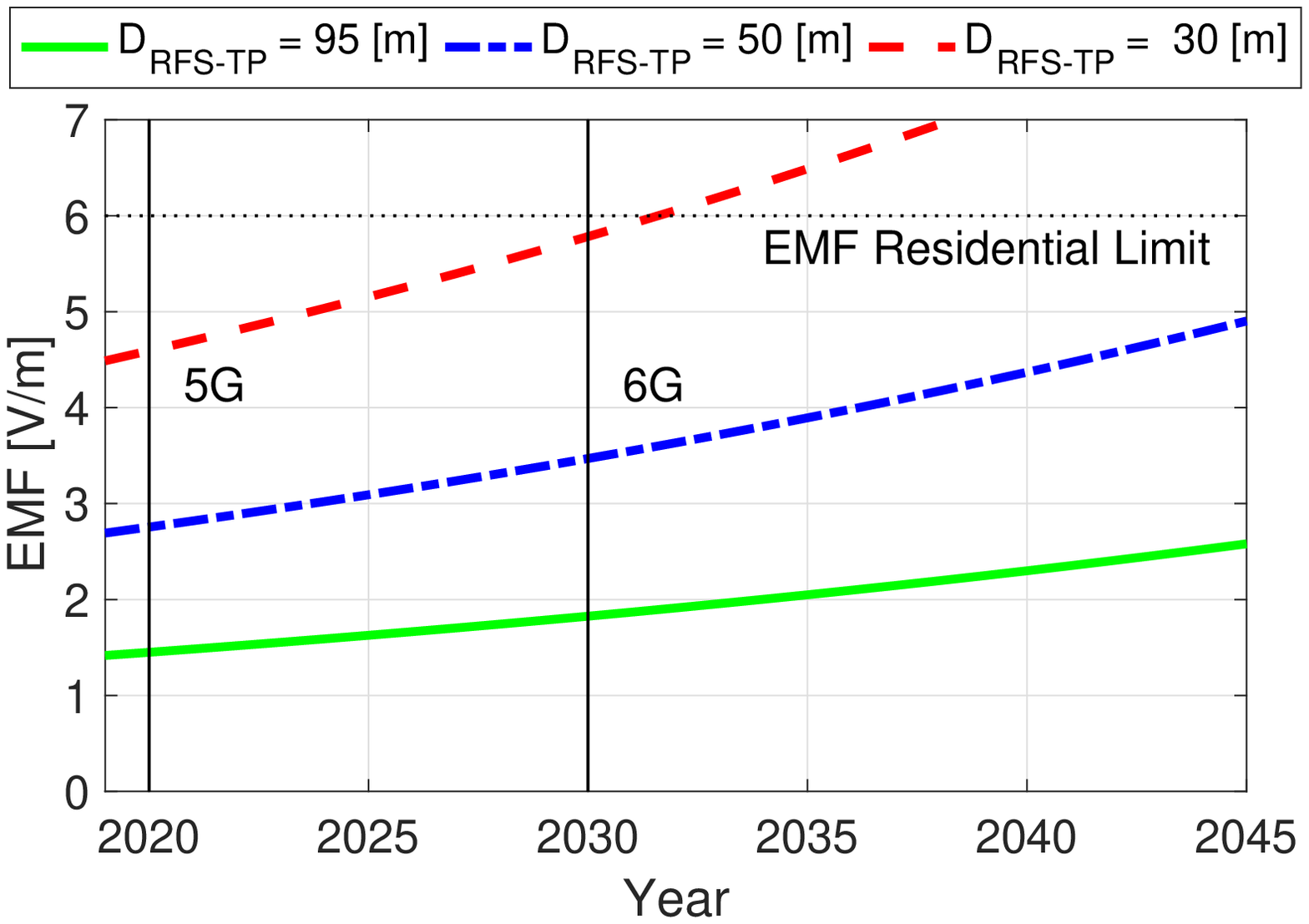}
    \label{fig:EIRP_prediction_var_distance_0_1}
}
\subfigure[$F_{\text{MIN}}$=0.5~V/m]
{
	\includegraphics[width=6.5cm]{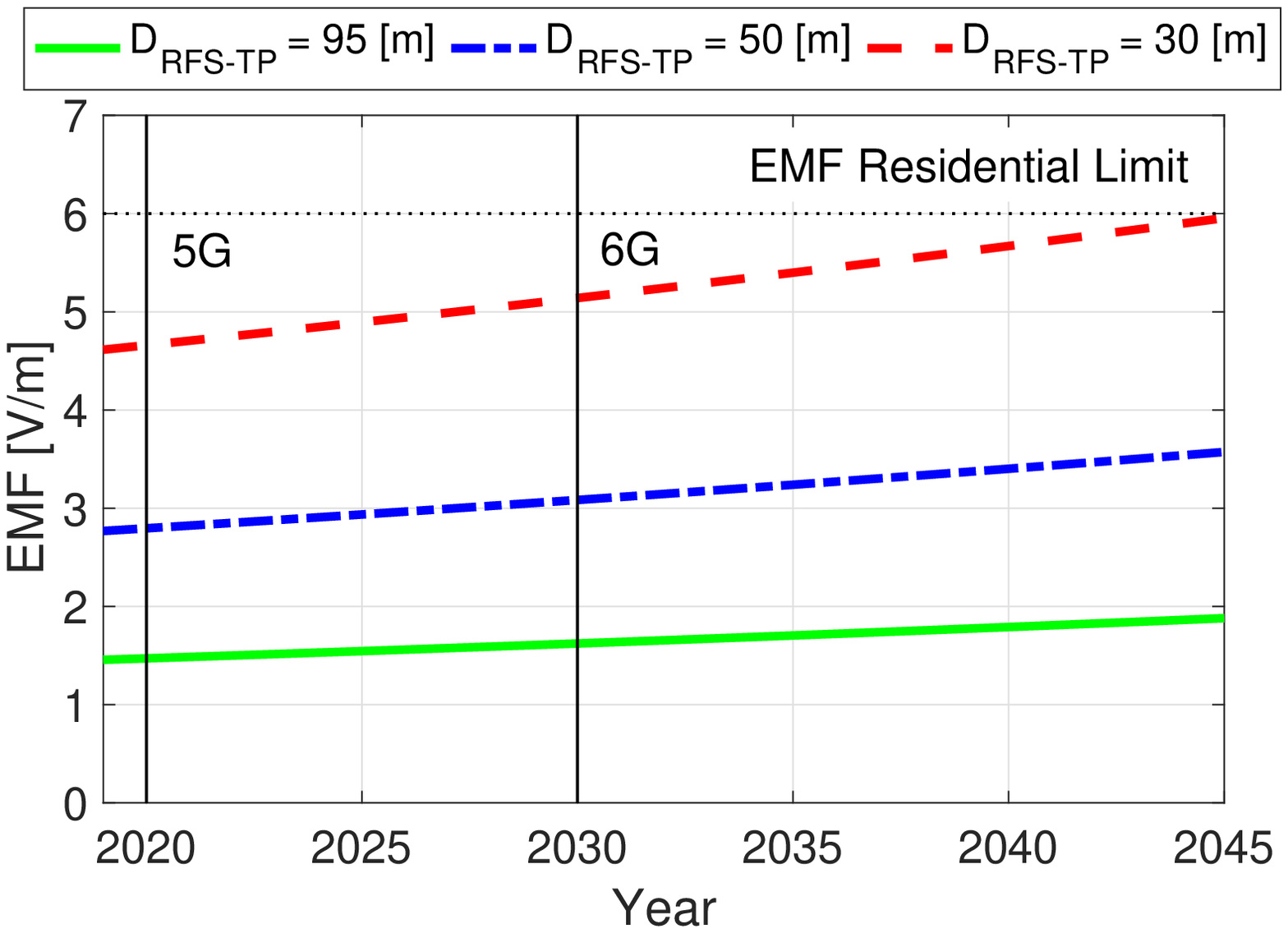}
	\label{fig:EIRP_prediction_var_distance_0_5}
}
\caption{EIRP-based prediction (B category).}
\label{fig:eirp_prediction_var_distance}
\vspace{-4mm}
\end{figure}

In the final part of our work, we adopt the very simple assumption that the future EMFs will continue to grow with the increasing trends which were observed for the EIRP during the past years. Although we recognize that future technology changes may impact this estimation, we provide here a set of (preliminary) results. Specifically, we initially compute the predicted EIRP values from Eq.~(\ref{eq:single:exp}) with parameters in Tab.~\ref{tab:estimated_parameters_EIRP} and year $y \in (2019,2045]$. We then compute the predicted EMF as: 
\begin{equation}
F^{\text{EST}}_y = \sqrt{\frac{G \cdot Z \cdot E_y}{4 \pi \cdot{D_{RFS-TP}}^{(2)}}}
\end{equation}
where $E_y$ is the predicted EIRP over $y \in (2019,2045])$, the values of $G$ and $Z$ are the ones used in Eq.~(\ref{eq:eirp_definition}), $E_y$ is the estimated EIRP at year $y$ (computed with Eq.~(\ref{eq:single:exp}), and the $D_{\text{RFS-TP}}$ is the RFS-TP distance, which we vary according to different values. Fig.~\ref{fig:EIRP_prediction_var_distance_0_1}-\ref{fig:EIRP_prediction_var_distance_0_5} report the obtained results, for $F_{\text{MIN}}=0.1$~[V/m] and $F_{\text{MIN}}=0.5$~[V/m], respectively. Moreover, we consider a variation $D_{\text{RFS-TP}}$ in the range $30-95$~[m]. Interestingly, the predicted EMF is going to notably increase over the years. This is due to the fact that the predicted EIRP will also increase, with a larger increase when assuming $F_{\text{MIN}}=0.1$~[V/m]. Moreover, when considering $D_{\text{RFS-TP}}$=30~[m], the 6~[V/m] residential limit will be saturated by 2031 and 2045 with $F_{\text{MIN}}=0.1$~[V/m] and $F_{\text{MIN}}=0.5$~[V/m] respectively. In the worst case, these saturation levels will prevent operators to install newer generation technologies (like advanced 5G and/or 6G networks, which will appear in the next decades). On the other hand, the fear of the population about BS exposure will increase, due to EMF levels close to the maximum limits.

\section{Summary and Future Works}
\label{sec:conclusions}

We have analyzed the EMF exposure in the vicinity of cellular towers over a large dataset covering 17 years of measurements. Our results point out several aspects. First, the EMF exposure from BSs is largely below the EMF limit enforced for residential areas, thus providing an adequate safety level for the population. Second, the EMF exposure tends to be notably reduced (i.e., less than 1~[V/m]) when the RFS-TP distance is more than 100~[m]. Third, BSs represent the lowest exposure compared to Radio/TV and other RFSs. Fourth, the measured EMF levels have been slightly increased over the year. This increase is likely due to the deployment of subsequent BS deployment, as well as the introduction of the $\alpha_{24}$ parameter. This observation is also corroborated by the EIRP-based analysis. As a side effect, the future EMF levels will (likely) reach the residential limit by the next twenty years at a distance of 30~[m] from the BSs. 

As future work, we plan to update our analysis as soon as 5G BSs will be fully deployed over the territory. To this aim, we will consider the impact of regulations on the exposure levels for people living in proximity to 5G BSs. Additionally, we plan to study the difference in exposure levels between the different areas of the territory, as well as a better assessment of indoor vs. outdoor EMF measurements. Finally, we plan to integrate more complex metrics, like integral-based ones \cite{vsuka2019application}, in our analysis.

\section*{Acknowledgements}
This work has received funding from the H2020 Locus Project (grant agreement n. 871249).

\bibliographystyle{ieeetr}

\end{document}